\documentclass[journal, 10pt]{IEEEtran}

\usepackage{graphicx}
\usepackage{subfig}
\usepackage{amsmath}
\usepackage{comment}

\usepackage{setspace}
\usepackage{url}
\usepackage{fancyvrb}

\date{}

\usepackage{enumitem}

\usepackage{url}

\usepackage{hyperref}

\ifCLASSINFOpdf
\else
\fi

\begin{document}

\onecolumn 

\begin{description}[style=multiline,leftmargin=3cm,font=\normalfont]

\item[\textbf{Citation}]{Y. Hu, Z. Long, and G. AlRegib, ``A High-Speed, Real-Time Vision System for Texture Tracking and Thread Counting,'' IEEE Signal Processing Letters, vol. 25, no. 6, pp. 758-762, 2018.}

\item[\textbf{DOI}]{\url{https://doi.org/10.1109/LSP.2018.2825309}}

\item[\textbf{Review}]{Date of publication: 9 March 2018}

\item[\textbf{Codes}]{\url{https://ghassanalregibdotcom.files.wordpress.com/2018/09/yuting_spl2018_code.zip}}

\item[\textbf{Bib}] {@article{hu2018high,\\
  title={A High-Speed, Real-Time Vision System for Texture Tracking and Thread Counting},\\
  author={Hu, Yuting and Long, Zhiling and AlRegib, Ghassan},\\
  journal={IEEE Signal Processing Letters},\\
  volume={25},\\
  number={6},
  pages={758--762},\\
  year={2018},\\
  publisher={IEEE}
}

}


\item[\textbf{Copyright}]{\textcopyright 2018 IEEE. Personal use of this material is permitted. Permission from IEEE must be obtained for all other uses, in any current or future media, including reprinting/republishing this material for advertising or promotional purposes,
creating new collective works, for resale or redistribution to servers or lists, or reuse of any copyrighted component
of this work in other works. }

\item[\textbf{Contact}]{\href{mailto:huyuting@gatech.edu}{huyuting@gatech.edu}  OR \href{mailto:zhiling.long@ece.gatech.edu}{zhiling.long@ece.gatech.edu} OR \href{mailto:alregib@gatech.edu}{alregib@gatech.edu}\\
    \url{http://ghassanalregib.com/} \\ }
\end{description}

\thispagestyle{empty}
\newpage
\clearpage
\setcounter{page}{1}

\twocolumn

\title{A high-speed, real-time vision system for texture tracking and thread counting}

\author{Yuting~Hu,~\IEEEmembership{Student Member,~IEEE,}
        Zhiling~Long,~\IEEEmembership{Member,~IEEE,}
        and~Ghassan~AlRegib,~\IEEEmembership{Senior Member,~IEEE} }

\markboth{IEEE SIGNAL PROCESSING LETTERS,~Vol.~25, No.~6, -~2018}%
{Shell \MakeLowercase{\textit{et al.}}: Bare Demo of IEEEtran.cls for IEEE Journals}

\maketitle

\begin{abstract}
In garment manufacturing, an automatic sewing machine is desirable to reduce cost. To accomplish this, a high-speed vision system is required to track fabric motions and recognize repetitive weave patterns with high accuracy, from a micro perspective near a sewing zone. In this paper, we present an innovative framework for real-time texture tracking and weave pattern recognition. Our framework includes a module for motion estimation using blob detection and feature matching. It also includes a module for lattice detection to facilitate the weave pattern recognition. Our lattice-detection algorithm utilizes blob detection and template matching to assess pair-wise similarity in blobs' appearance. In addition, it extracts information of dominant orientations to obtain a global constraint in the topology. By incorporating both constraints in the appearance similarity and the global topology, the algorithm determines a lattice that characterizes the topological structure of the repetitive weave pattern, thus allowing for thread counting. In our experiments, the proposed thread-based texture tracking system is capable of tracking denim fabric with high accuracy (e.g., 0.03$^{\circ}$ rotation and 0.02 weave-thread' translation errors) and high speed (3 frames per second), demonstrating its high potential for automatic real-time textile manufacturing.
\end{abstract}

\begin{IEEEkeywords}
Texture tracking, lattice detection, thread counting, texture regularity, pattern recognition.
\end{IEEEkeywords}

\IEEEpeerreviewmaketitle

\vspace{-0.2in}

\section{Introduction}

\begin{figure}[t]
    \centering
    \begin{minipage}[!htbp]{0.2\linewidth}
      \centering
      \centerline{\includegraphics[width=5cm]{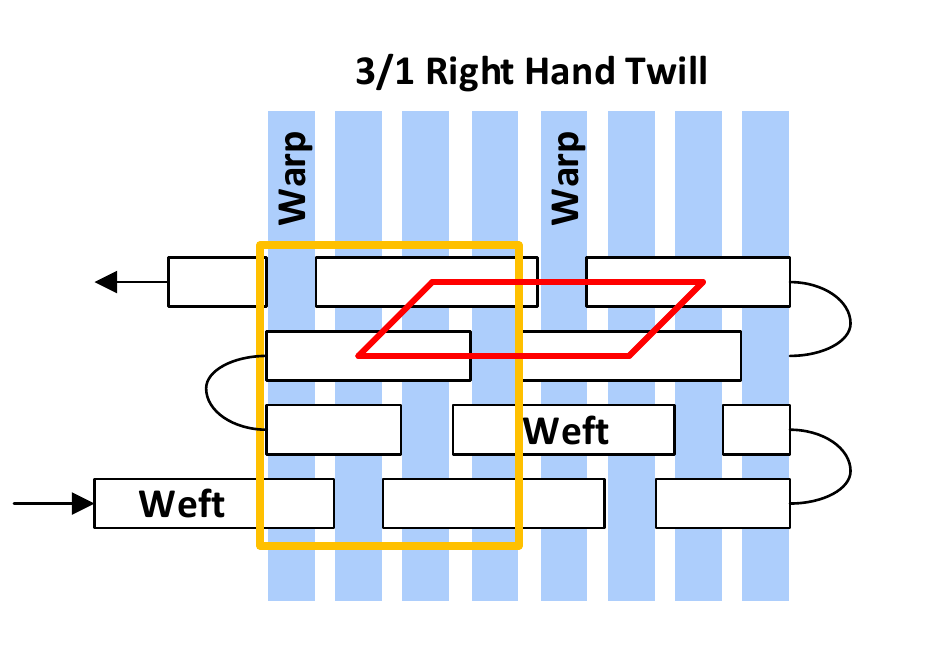}}
    \end{minipage}
\caption{An example of twill weave pattern}
\label{fig:twill}
\end{figure}

\IEEEPARstart{T}{extures} are ubiquitous in real world, and texture analysis has been studied for years in the field of image processing and computer vision. One application in textile industry is an automatic sewing machine for garment manufacturing\footnote{This work was partially supported by a Walmart Foundation grant ($\#$1806K45).}, which requires a vision system to recognize fabric patterns and track fabric motions. Although texture tracking in general has been investigated in the literature~\cite{wang2015noise,pham2017sar}, few work focuses on robust and efficient algorithms specially designed for fabrics in a real-time environment. In the cutting or sewing part of the sewing machine, to avoid aliasing, a real-time vision system should be capable of continuously monitoring a small region of a fabric near the sewing needle and tracking a very small motion (possibly less than the width of a thread) in successive frames. Besides fabric tracking, the vision system should be capable of thread counting, which reduces the effect of the local fabric distortions on tracking accuracy. To illustrate the concepts of weave patterns, threads, and lattices, we show an example in Fig.~\ref{fig:twill}. Given an original starting point and a center point in a thread-based coordinate system, counting threads denotes maintaining a cumulative amount of warp and weft (filling) thread that has passed the center point. 

To design a fast and robust vision system for an automatic sewing machine, Book et al.~\cite{book2010automated} proposed a prototype in which the concept of thread-count was first introduced. In the vision system, they used the Harris corner detection algorithm to detect corner features and fabric translation, and 2D-fast Fourier transform (FFT) to track fabric angles. However, the FFT-based algorithm does not work well for accurately estimating a small rotation angle (e.g. 0.1$^\circ$). In addition, thread-count remains a concept, with no solution provided in their work. Recognizing fabric threads is closely related to discovering texture regularity or granularity. In the literature, Liang and Weller~\cite{liang2016edge} detected granularity (i.e., the size of texture primitives) for general texture with simple edge detection techniques. To discover the lattices of near-regular texture (NRT), Hays et al.~\cite{hays2006discovering} first formulated a lattice-finding problem for NRT as a higher-order correspondence problem. This technique uses interest point detectors, iteratively proposes and assigns neighboring texture primitives, and then seeks an optimal lattice assignment by maximizing the pair-wise visual similarity and the geometric consistency. The approach fits well for thread detection and counting, however, the optimization and iteration process is complex and time consuming, not suitable for real time applications.
In addition, Lin and Liu~\cite{lin2006tracking,lin2007lattice} proposed the first deformed lattice detection and tracking algorithm for dynamic NRTs and compared it with optical flow and Lukas-Kanade algorithms~\cite{lucas1981iterative}. Their methods (as well as~\cite{liu2015patchmatch}) are also computationally expensive by involving MRF-based tracking models.

In this paper, we propose an efficient, robust, and accurate feature-based approach to track individual fabric threads and provide the associated motion information in terms of position and orientation. Our contributions include:
\begin{enumerate}
    \item an innovative framework that integrates a lattice detection module to accomplish fabric tracking in a thread-based coordinate system instead of a pixel-based system, to ensure robustness to local fabric deformation;
    \item a novel algorithm for fast and efficient lattice detection for thread counting, achieved by constraining on both the appearance similarity (through template learning and matching) and the global topology; and
    \item an extensive comparative study evaluating various methods of keypoint detection and description for their applicability to the fabric tracking problem of interest.
\end{enumerate}

\begin{figure}[t]
    \centering
    \begin{minipage}[!htbp]{0.9\linewidth}
      \centering
      \centerline{\includegraphics[width=9cm]{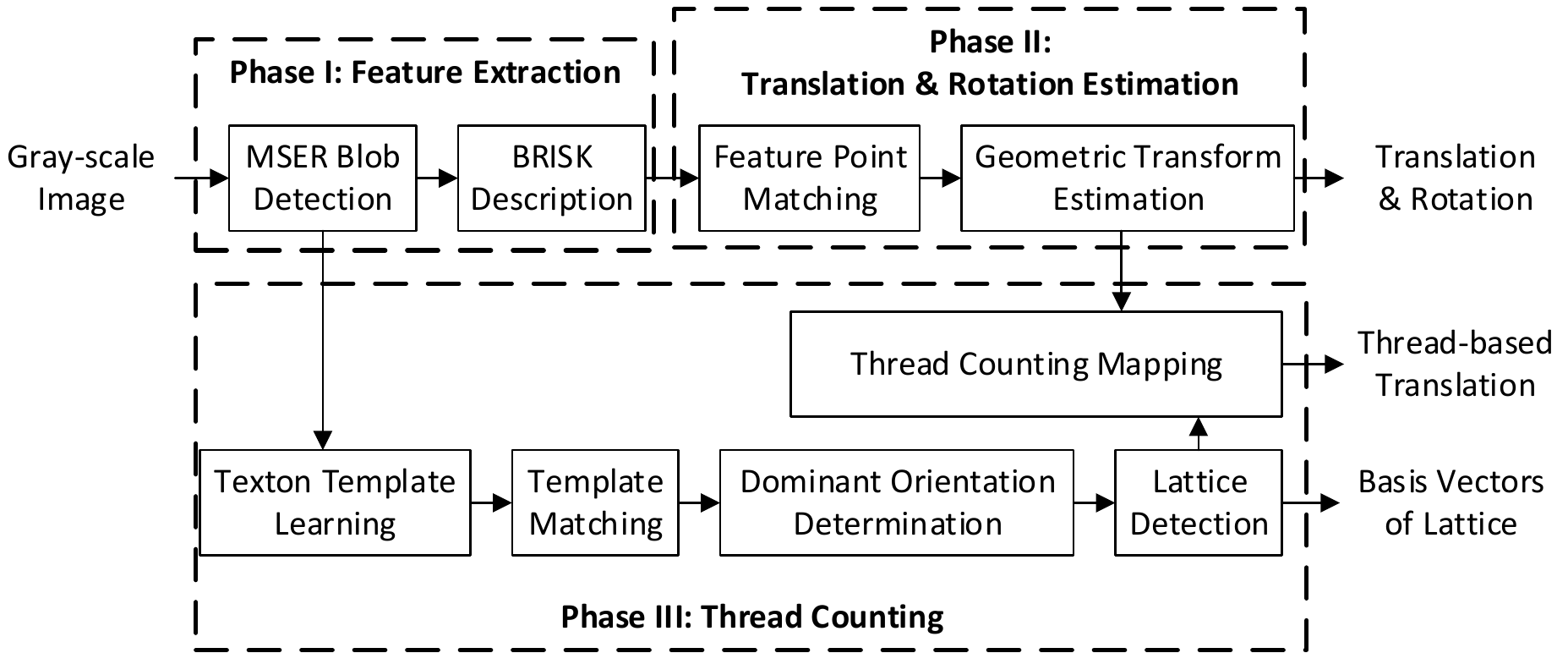}}
    \end{minipage}
\caption{The flowchart of the proposed framework}
\label{fig:flowchart}
\end{figure}

\vspace{-0.2in}

\section{Texture Tracking and Thread Counting}
\label{sec:feature-based texture tracking framework}
\begin{figure}[!t]
\centering
\subfloat[Raw texture image]{\includegraphics[width=2.6cm]{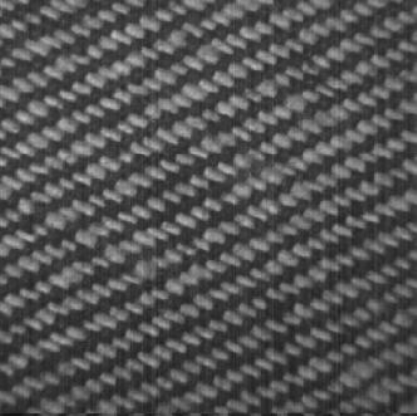}%
\label{fig:featureextraction_rawimage}}
\hfil
\subfloat[MSER feature detection]{\includegraphics[width=2.6cm]{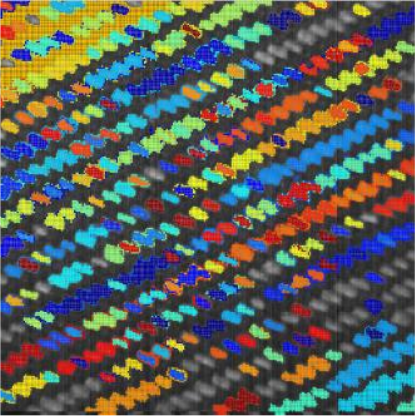}%
\label{fig:featureextraction_colorMSER}}
\hfil
\subfloat[BRISK feature description]{\includegraphics[width=6.4cm]{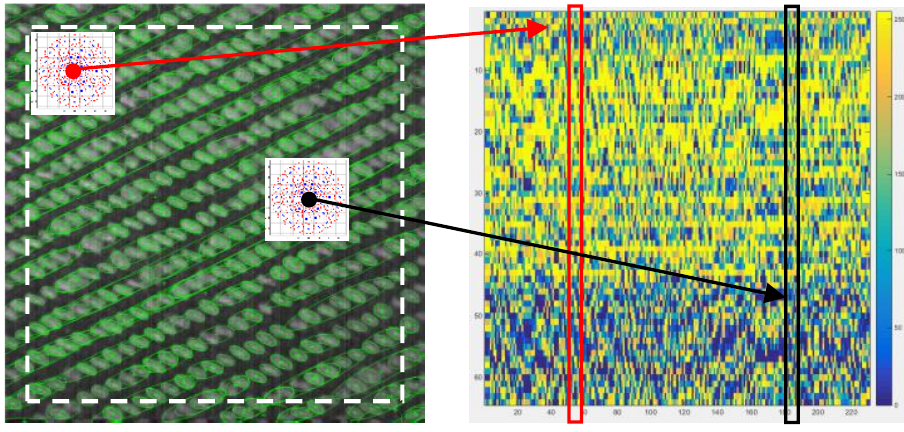}%
\label{fig:featureextraction_BRISK}}
\caption{Feature extraction}
\label{fig:featureextraction}
\end{figure}

We present a framework to automatically track small motions and detect the lattice in near-regular texture (e.g. denim fabric shown in Fig.~\ref{fig:featureextraction_rawimage}).  The framework includes three phases shown in Fig.~\ref{fig:flowchart}. Phase I is feature extraction in which we detect feature points and describe the local region around each feature point. In phase II, we find matching feature points between two sets of feature points in two successive frames and estimate a geometric transformation with translation and rotation offsets. In phase III, we learn the repetitive pattern template and seek a local lattice by involving both local appearance similarities and underlying topological relationship among texture patterns.

\vspace{-0.15in}
\subsection{Feature Detection and Description}
\label{ssec:featureextraction}

To establish reliable matching between two frames, we need a feature detector that  extracts ample feature points and a feature descriptor that distinctively describes local regions.

\subsubsection{MSER feature detection}
 Since the image exhibits abundant bright blob regions with a near-regular placement and that a large percentage of the same blobs appear in successive frames, we choose maximally stable extremal region (MSER)~\cite{donoser2006efficient} to detect stable blobs. The detected MSERs are visualized in Fig.~\ref{fig:featureextraction_colorMSER}, in which each unique color represents one individual MSER region. After MSER detection, we fit ellipses and centroids into detected regions displayed in green ellipses and points in Fig.~\ref{fig:featureextraction_BRISK}. We utilize MSER regions later to match feature points in two frames and to generate a texture template. \\

\vspace{-0.15in}
\subsubsection{BRISK feature description}
For feature description, binary robust invariant scalable keypoints (BRISK)~\cite{leutenegger2011brisk} offers a
fast alternative to the well-known algorithms (e.g. scale-invariant feature transform (SIFT) / speeded up robust features (SURF)), and still maintains comparable matching performance. We mix each MSER feature point with a BRISK descriptor.
Taking Fig.~\ref{fig:featureextraction_BRISK} as an example, we use the red column vector to represent the BRISK feature vector of the red MSER feature point. It is worth noting that, for feature description in this paper, using typical texture features (e.g.~\cite{cimpoi2016deep,hu2017sselbp,hu2016completed, liu2017local}) normally cannot satisfy both the robustness and the high-speed requirements simultaneously.

\vspace{-0.15in}

\subsection{Translation and Rotation Estimation}
We estimate translation and rotation offsets between two frames from feature points and feature vectors through feature-point matching and geometric transformation estimation.

\begin{figure}[t]
    \centering
    \begin{minipage}[!htbp]{0.8\linewidth}
      \centering
      \centerline{\includegraphics[width=7cm]{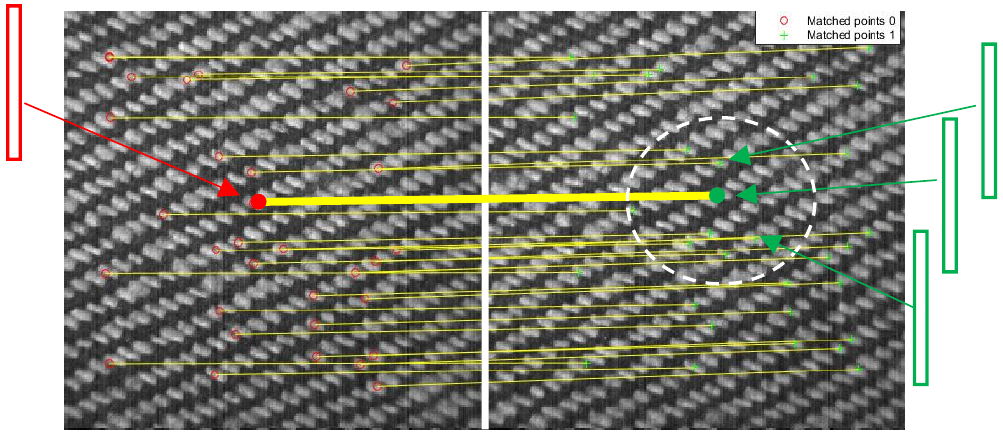}}
    \end{minipage}
\caption{Feature point matching}
\label{fig:translationandrotationestimation}
\end{figure}

\subsubsection{Feature-point matching}
Feature-point matching involves finding corresponding interest points between a pair of images. Since BRISK feature descriptors are binary strings, we use the Hamming distance for computational efficiency. We apply the nearest neighbor approach, in which a match threshold is set for selecting the strongest matches. Therefore, using local neighborhood information given by feature descriptors, we find reliable matching points. Fig.~\ref{fig:translationandrotationestimation} illustrates feature-point matching, in which we use matched pairs to estimate the geometric transformation between a pair of images.

\subsubsection{Geometric transformation estimation}
\label{sec:affinematrix}

In an affine matrix
\begin{math}
\begin{bmatrix}
     cos\bigtriangleup \theta & -sin\bigtriangleup \theta & 0 \\
     sin\bigtriangleup \theta & cos\bigtriangleup \theta & 0 \\
     \bigtriangleup x& \bigtriangleup y & 1
    \end{bmatrix}
\end{math}, $\bigtriangleup x$ and $\bigtriangleup y$ note translation offsets in camera pixels and $\bigtriangleup \theta$~($^{\circ}$) notes a rotation angle. To estimate the parameters of a mathematical model from a set of observations that contain outliers, we use the m-estimator sample consensus (MSAC)~\cite{torr2000mlesac}, one variant of random sample consensus (RANSAC)~\cite{fischler1981random}, to estimate a 2-D geometric transform from matching pairs, which obtains a global affine matrix in a standard orthogonal coordinate system and reports translation and rotation offsets.

\vspace{-0.15in}

\subsection{Thread Counting}

Thread counting starts with low-level vision cues (e.g. blobs) and ends with high-level lattice models shown in Fig.~\ref{fig:threadcounting}. We generate a representative blob template and seek a vector pair consistent with geometric relationships between blobs.

\subsubsection{Template learning}
As shown in Fig.~\ref{fig:featureextraction}(b), MSER generates potential blobs, some of which blobs are connected, and others are not. From attributes of each blob region (e.g. its area and its intensity values), we group blobs into two clusters: individual blobs and grouping blobs. We use all individual blobs such as the highlighted blob regions shown in Fig.~\ref{fig:threadcounting_blobcandidates} to propose a blob template. We align all individual blobs according to their centroids and average their intensity values to determine a blob template shown in Fig.~\ref{fig:threadcounting_blobtemplate}.

\subsubsection{Template matching}
We use the obtained blob template to detect the nearest neighboring blobs of the current MSER center (i.e., the red point in Fig.~\ref{fig:threadcounting_templatematching}). To find neighboring blobs, we adopt the correlation-based template matching method, which utilizes the information in local peaks on a correlation map between the candidate neighboring blob region and the blob template.
In Fig.~\ref{fig:threadcounting_templatematching}, we show the centroids of detected neighboring blobs in blue and use them as a constraint of appearance similarities for later lattice detection.

\subsubsection{Dominant orientation determination}
\begin{figure}[!t]
\centering
\subfloat[Blob selection]{\includegraphics[width=2.7cm]{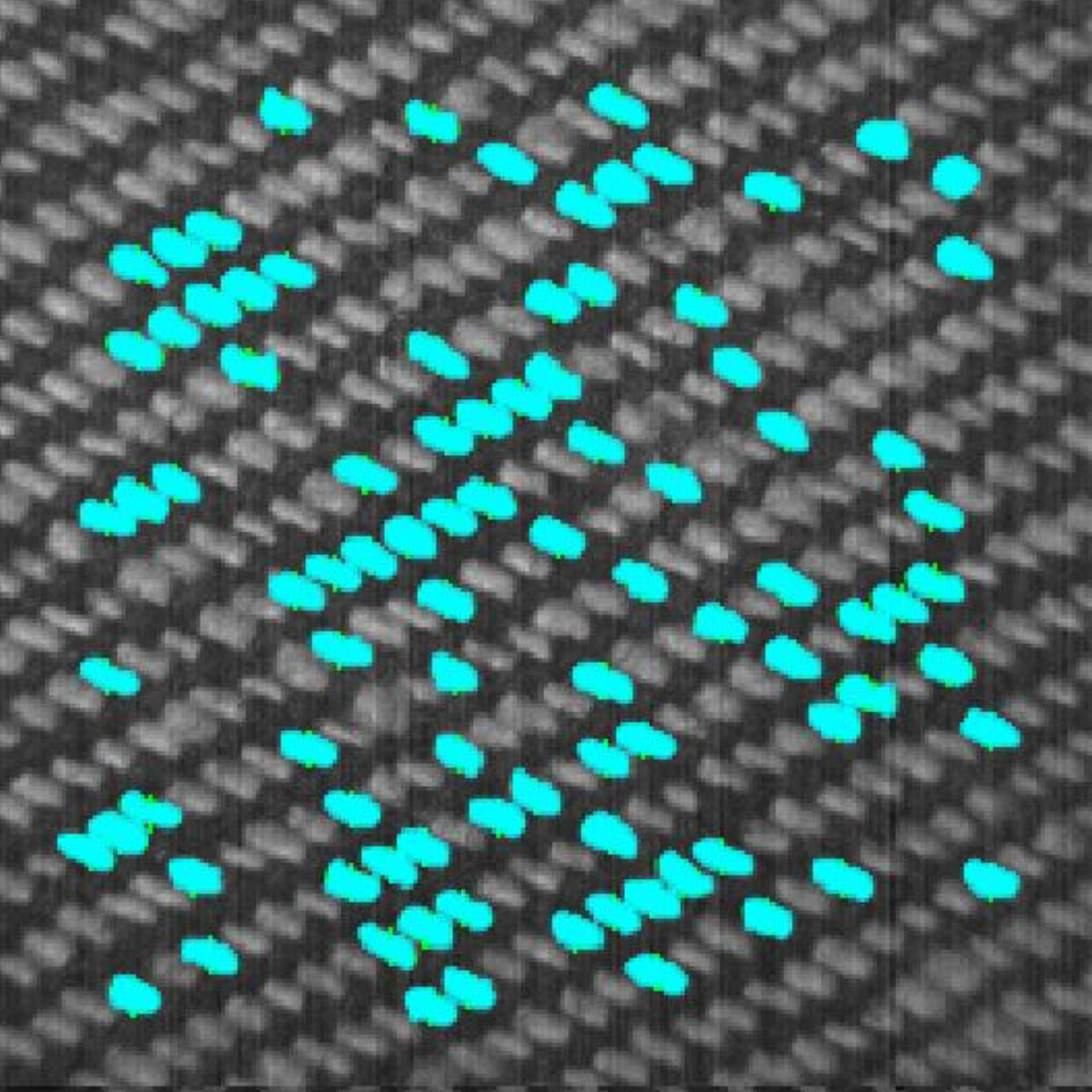}
\label{fig:threadcounting_blobcandidates}}
\hfil
\subfloat[Blob template]{\includegraphics[width=2.7cm]{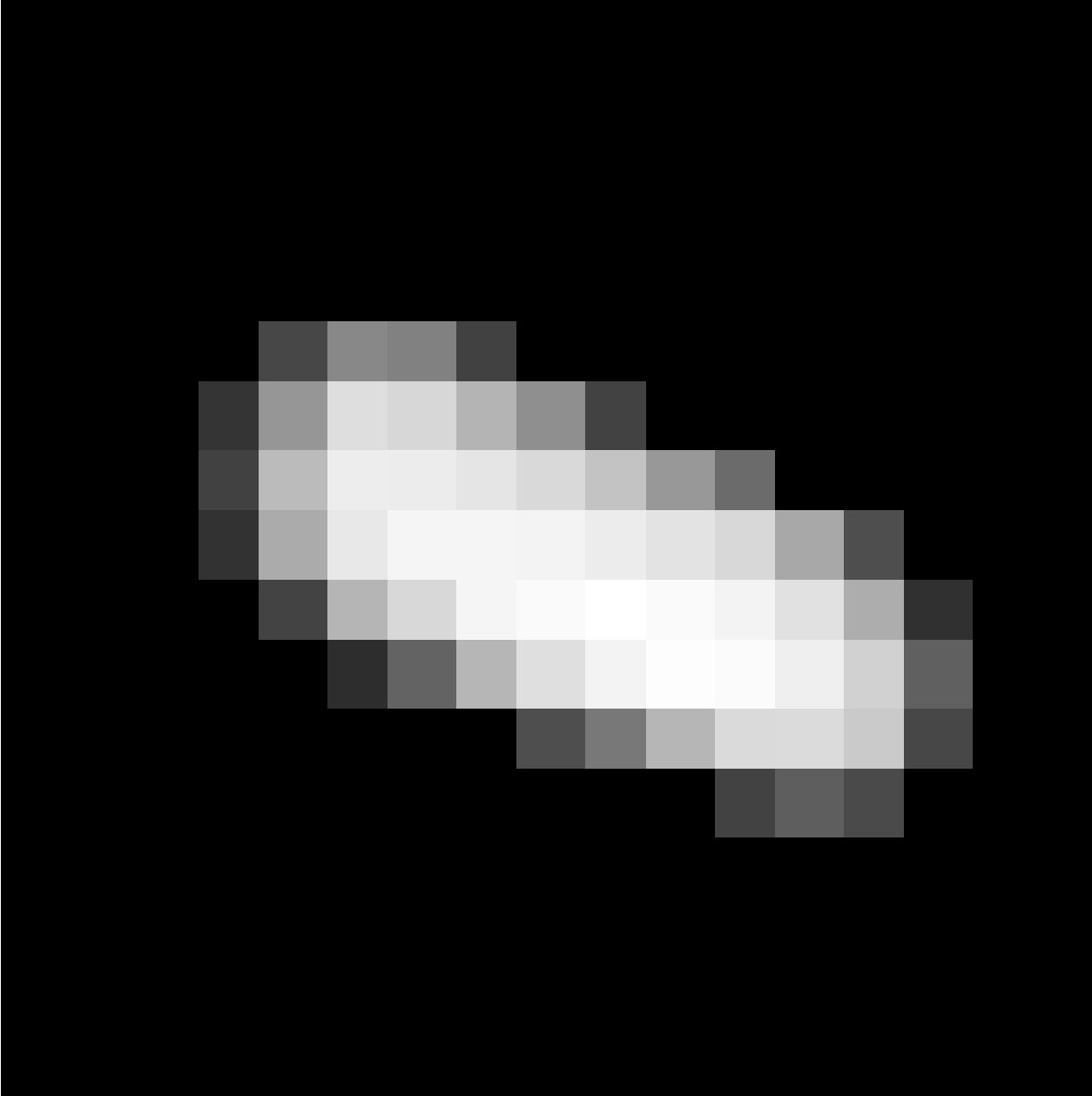}%
\label{fig:threadcounting_blobtemplate}}
\hfil
\subfloat[Template matching]{\includegraphics[width=2.7cm]{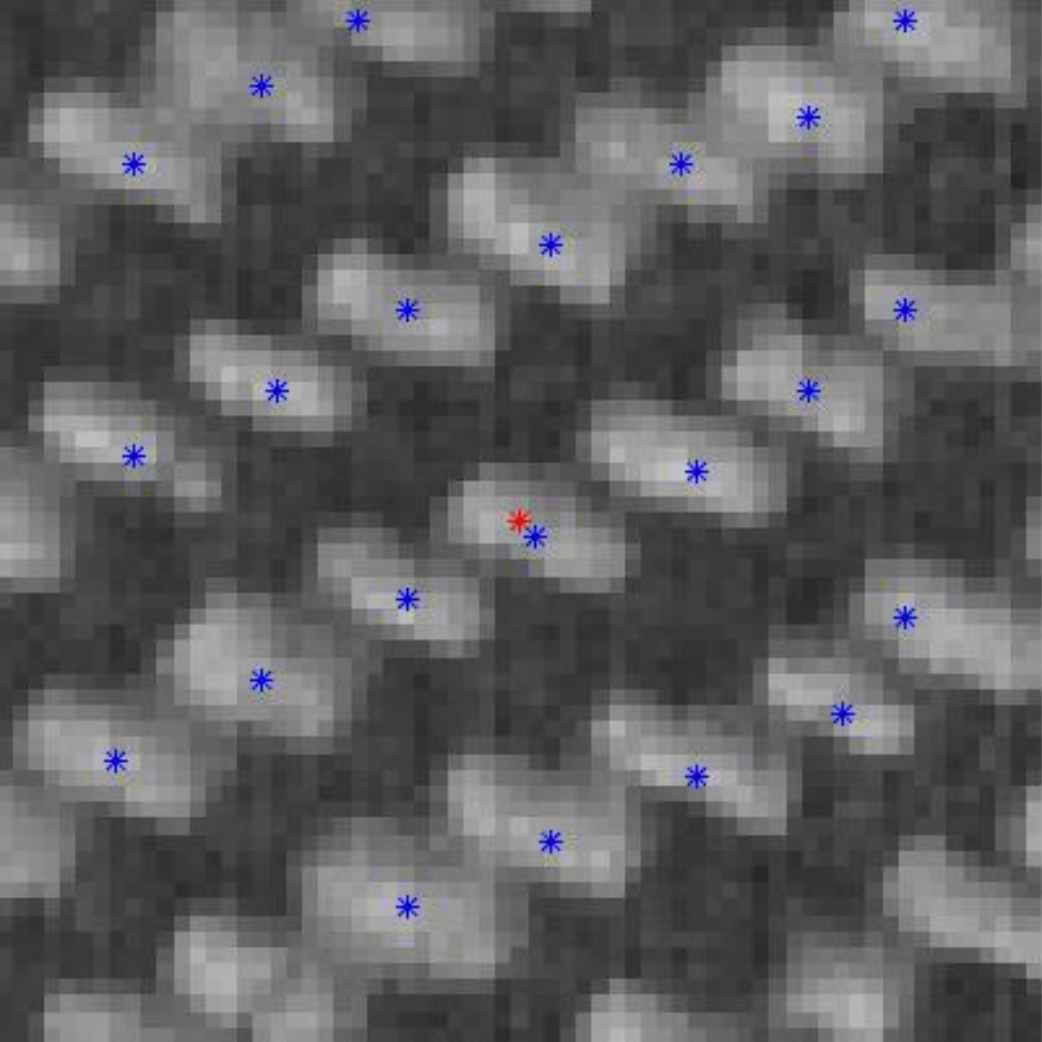}%
\label{fig:threadcounting_templatematching}}
\hfil
\subfloat[Frequency domain]{\includegraphics[width=2.7cm]{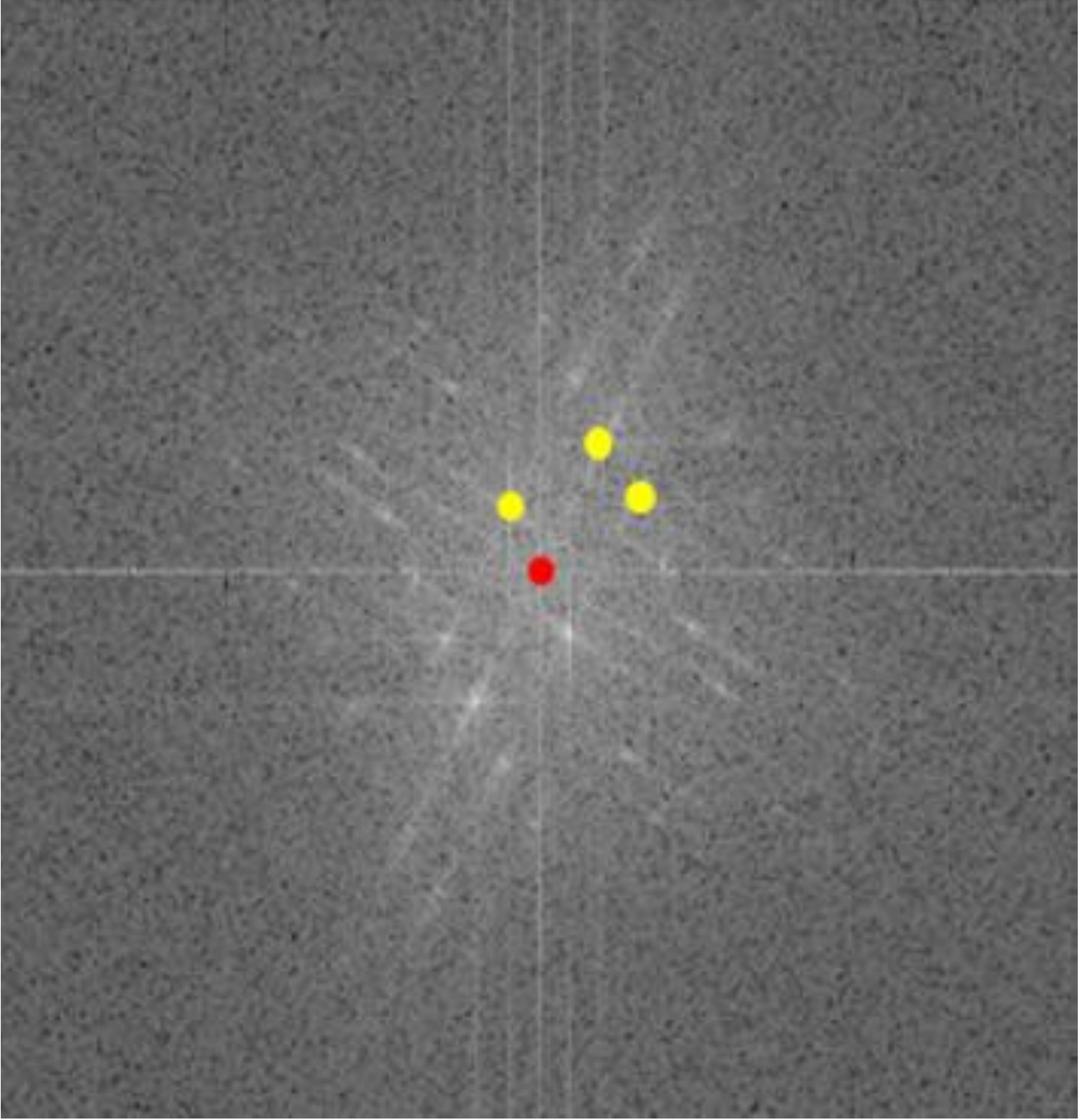}%
\label{fig:threadcounting_frequencydomain}}
\hfil
\subfloat[Dominant directions]{\includegraphics[width=2.7cm]{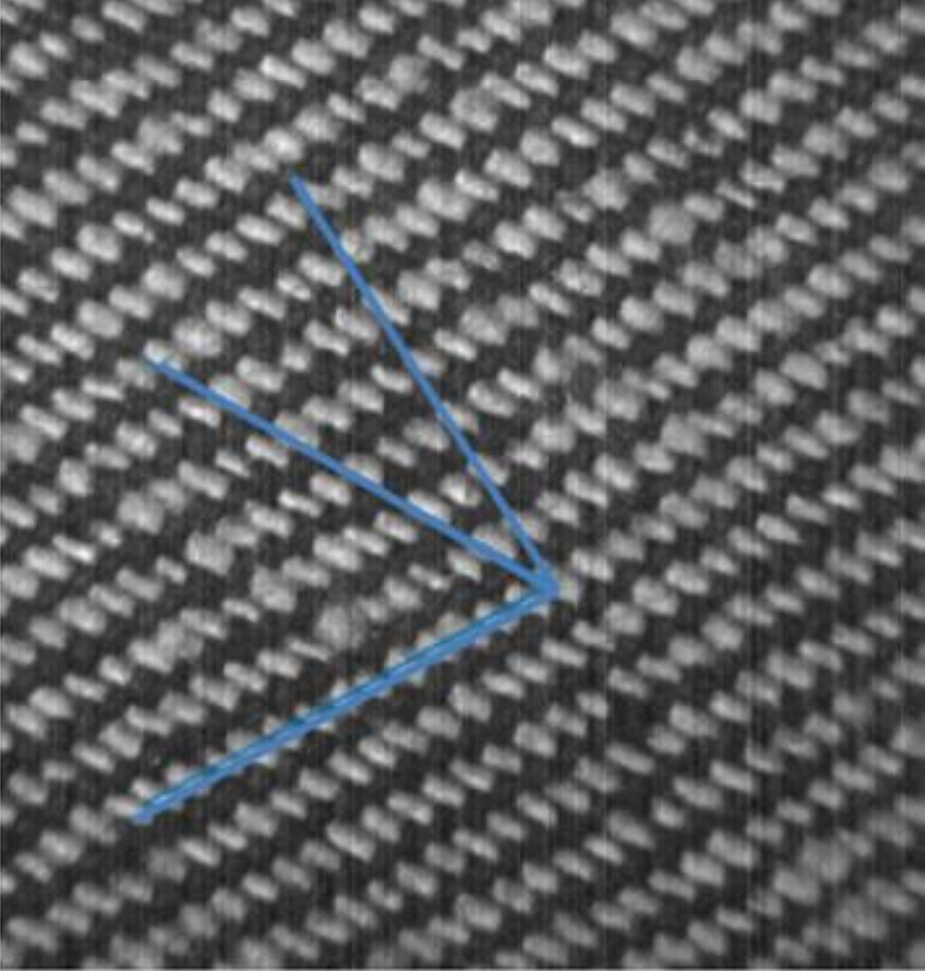}%
\label{fig:threadcounting_dominantdirection}}
\hfil
\subfloat[Lattice detection]{\includegraphics[width=2.8cm]{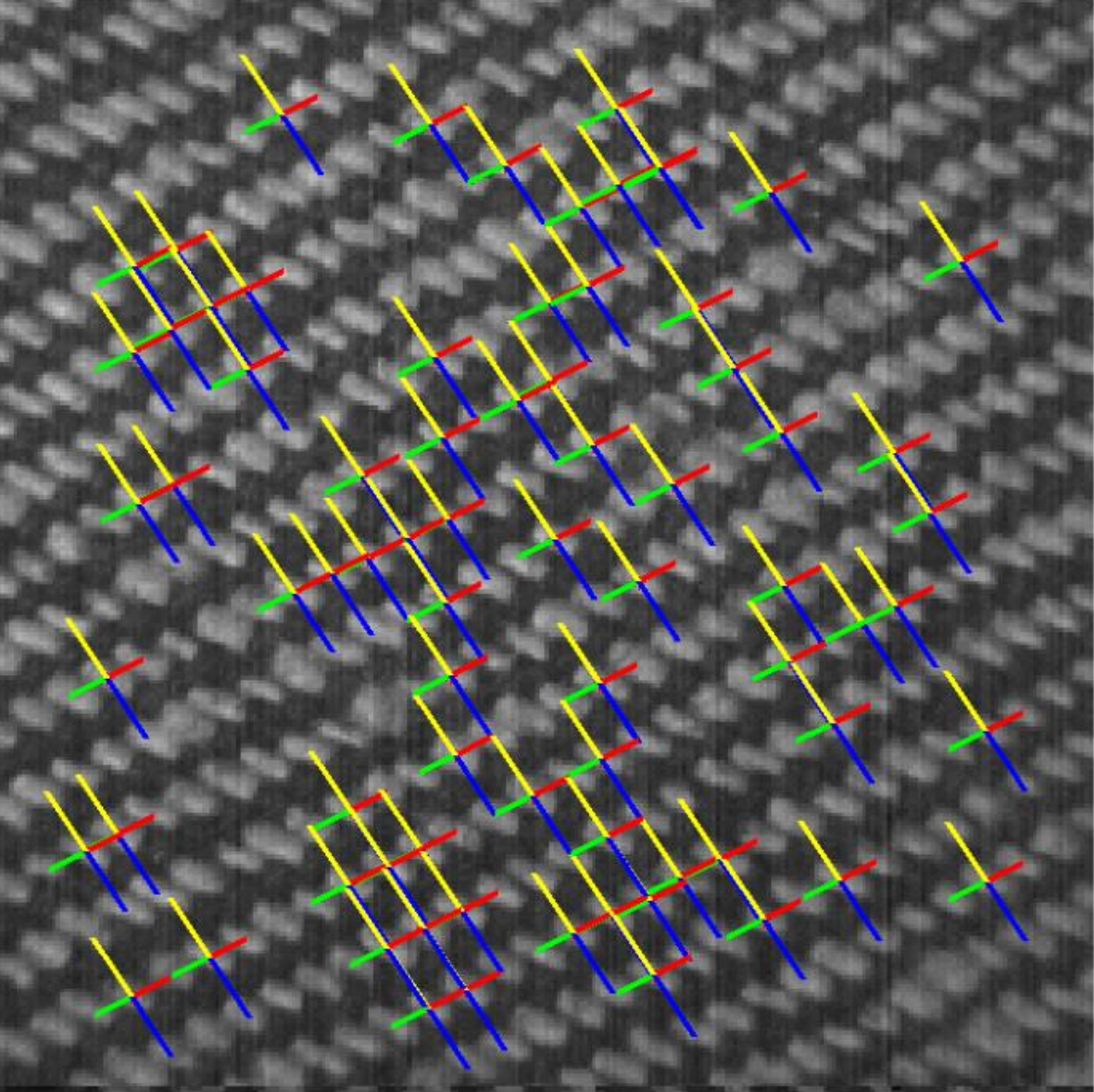}%
\label{fig:threadcounting_latticedetection}}
\caption{Thread counting}
\label{fig:threadcounting}
\end{figure}

 Blobs form a near-regular placement of repetitive patterns in dominant orientations. For a square image, the angular orientation of a peak AC component in the frequency domain and a dominant repeating orientation in the spatial domain are perpendicular. For example, the 2D-FFT of a denim image is shown in Fig.~\ref{fig:threadcounting_frequencydomain}, in which the red point represents a DC component; the yellow points represent AC components with strong peaks; and three AC peaks (yellow points) in the frequency domain correspond to three dominant orientations in the spatial domain (blue lines) in Fig.~\ref{fig:threadcounting_dominantdirection}. We use only two orientations from three as reference orientations. Peak features in the frequency domain help determine dominant directions and provide a geometric constraint for later lattice detection.

\subsubsection{Lattice detection}
Proposing a lattice model represents determining a vector pair connecting the current blob centroid and its two nearest matched neighbors. Orientations of the two basis vectors should follow the guidance of obtained dominant directions, which results in global topological consistency. For potential vector pairs, we minimize the distance defined in Eq.~\ref{eq:latticevector_distdef} and generate a final lattice proposal.
\begin{equation}
\label{eq:latticevector_distdef}
d(\vec{x},\vec{y})=\|\vec{x}-\vec{y}\|+w\|\theta_{\vec{x}-\vec{y}}-\theta_{ref}\|,
\end{equation}
where $\vec{x}$ and $\vec{y}$ represent coordinates of the current blob centroid and the candidate blob centroid, respectively; $\theta_{\vec{x}-\vec{y}}$ denotes the angle of the vector connecting the current blob centroid and the candidate blob centroid; $\theta_{ref}$ represents a dominant direction of repetitive patterns estimated from the frequency domain; $\|\vec{x}-\vec{y}\|$ denotes a constraint of appearance similarities from template matching; and $\|\theta_{\vec{x}-\vec{y}}-\theta_{ref}\|$ notes a topological constraint. To balance the appearance constraint and the topological constraint, we use $w$, a weighting factor related to prior knowledge of weave patterns. By selecting candidate blobs with the smallest $d(\vec{x},\vec{y})$, we determine basis vectors along the two dominant directions. We superimpose the final proposal of a local lattice on inlier MSER feature points shown in Fig.~\ref{fig:threadcounting_latticedetection}. We produce the final lattice proposal by involving both the similarity of the pair-wise texture appearance and the global consistency of topological relationships.

\subsubsection{Thread counting mapping}
To obtain fractional thread counting, thereby reducing the effect of fabric distortion for each inlier feature point, we calculate the translation vector and decompose it into the local lattice coordinate system with the assumption that we have a prior knowledge of the fabric type and mapping information between the local lattice and the physical fabric thread.

\vspace{-0.1in}

\section{Experiments and Discussion}
\label{experiments}
\begin{figure}[t]
    \centering
    \begin{minipage}[!htbp]{0.85\linewidth}  
      \centering
      \centerline{\includegraphics[width=8cm]{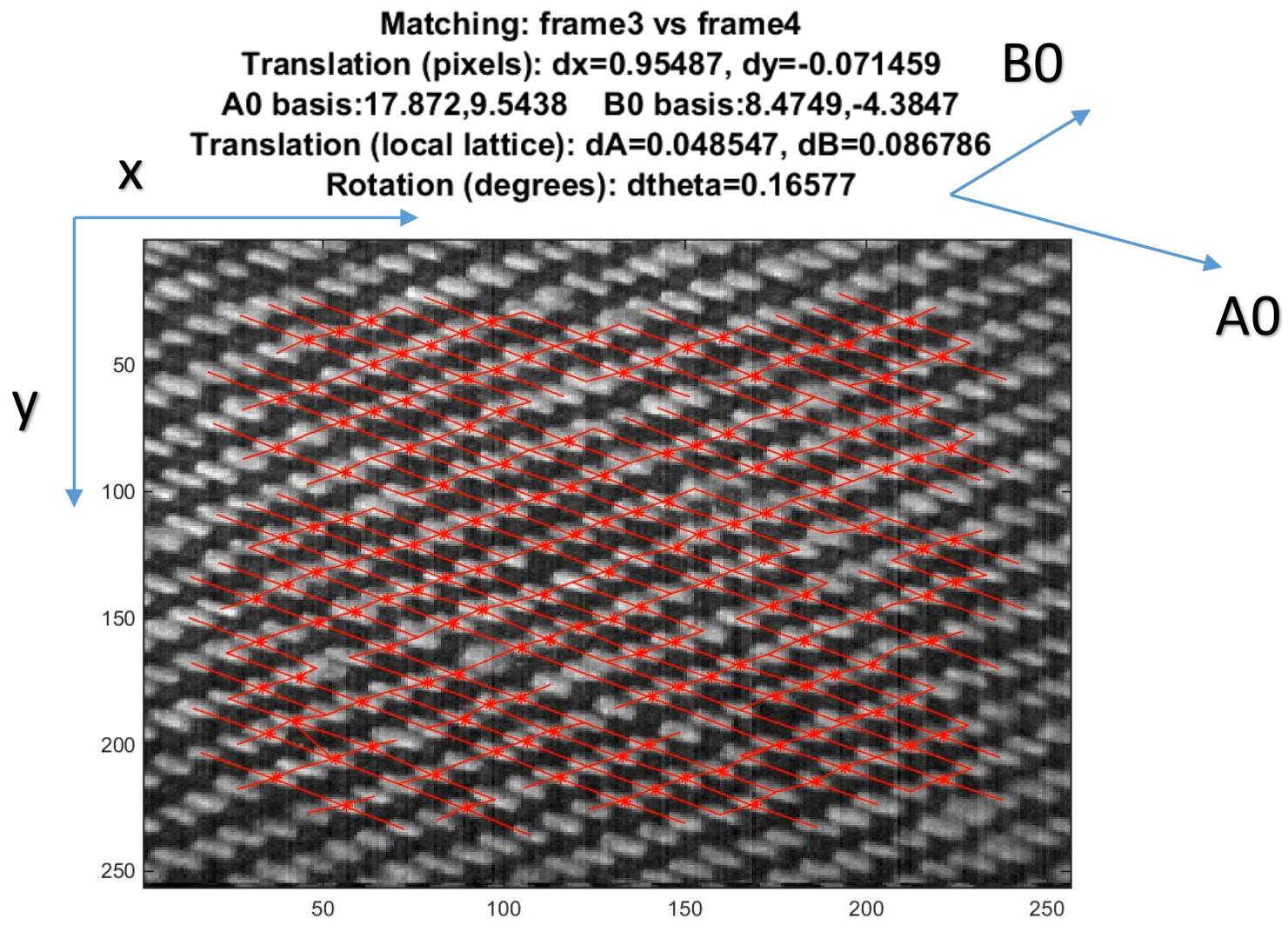}}
    \end{minipage}
\caption{A demo for real-time thread tracking.
}
\label{fig:HSVSdemo}
\end{figure}

To evaluate the tracking performance of our proposed framework, we conduct a set of experiments, in which we estimated a translation offset in a camera space, a rotation angle, and a translation offset in a thread-based coordinate system between two frames with small motions shown in Fig.~\ref{fig:HSVSdemo} (a demo video available online\footnote{\label{note1}https://ghassanalregib.com/texture-tracking-in-video-streams-and-weave-pattern-recognition/}). Our target texture is a piece of denim fabric mounted on a micro stage that allows precise translation and rotation. Since our camera captures only a small region of denim fabric, the field of view (FOV) of the captured images contains only the texture of denim fabric rather than the background texture. The size of the images is $256 \times 256$. We implement algorithms on MATLAB\textregistered 2014b with a PC (Intel i7-4790K, 4GHz, RAM: 32GB).


\begin{figure}[t]
    \centering
    \begin{minipage}[!htbp]{0.85\linewidth}
      \centering
      \centerline{\includegraphics[width=8.5cm]{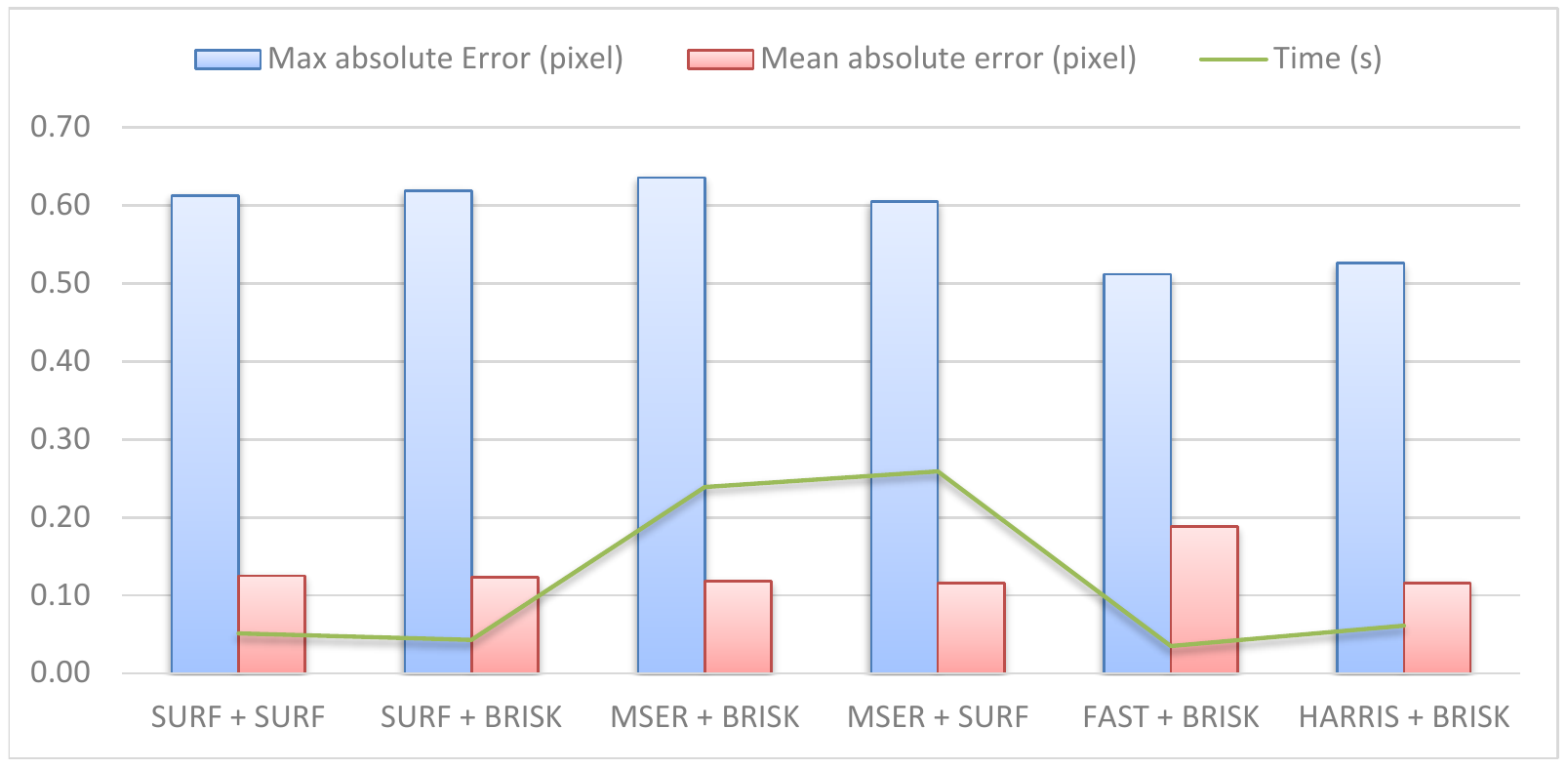}}
    \end{minipage}
\caption{Performance evaluation of translation estimation.}
\label{fig:results_translation}
\end{figure}

\textbf{\textit{Translation estimation}}: To evaluate the accuracy of translation estimation, we obtained ground truth using the micro stage and conducted an extensive experiment. We translated the micro stage from zero to ten mm at intervals of 0.5 mm (i.e., around 7.53 pixels) in the horizontal direction and acquired 20 test images for our experiment. With the ground truth of translation offsets, we combined various feature detectors and descriptors (e.g. SURF~\cite{bay2008speeded}, MSER~\cite{donoser2006efficient}, BRISK~\cite{leutenegger2011brisk}, FAST~\cite{rosten2005fusing}, and HARRIS~\cite{harris1988combined}) and compared their estimation accuracy. To quantify accuracy for translation estimation, we use three metrics: (1) the maximum value of the absolute error between the estimated and actual translation values in pixels; (2) the mean value of the absolute error in pixels; and (3) the computational cost in seconds. We demonstrate the performance of the six algorithms on tracking translation in Fig.~\ref{fig:results_translation}, in which ``A+B'' denotes ``feature detector+feature descriptor.'' From Fig.~\ref{fig:results_translation}, we observe: (1) ``MSER+SURF'' yields the lowest mean absolute error (i.e., 0.12 pixels); and (2) ``FAST+BRISK'' generates the lowest maximum absolute error (i.e., 0.51 pixels) and the lowest computational time (i.e., 0.035 seconds). To determine which algorithm to use, besides the three metrics mentioned above, we evaluate the accuracy and computation time of rotation estimation and thread counting.


\begin{figure}[t]
    \centering
    \begin{minipage}[!htbp]{0.85\linewidth}
      \centering
      \centerline{\includegraphics[width=8.5cm]{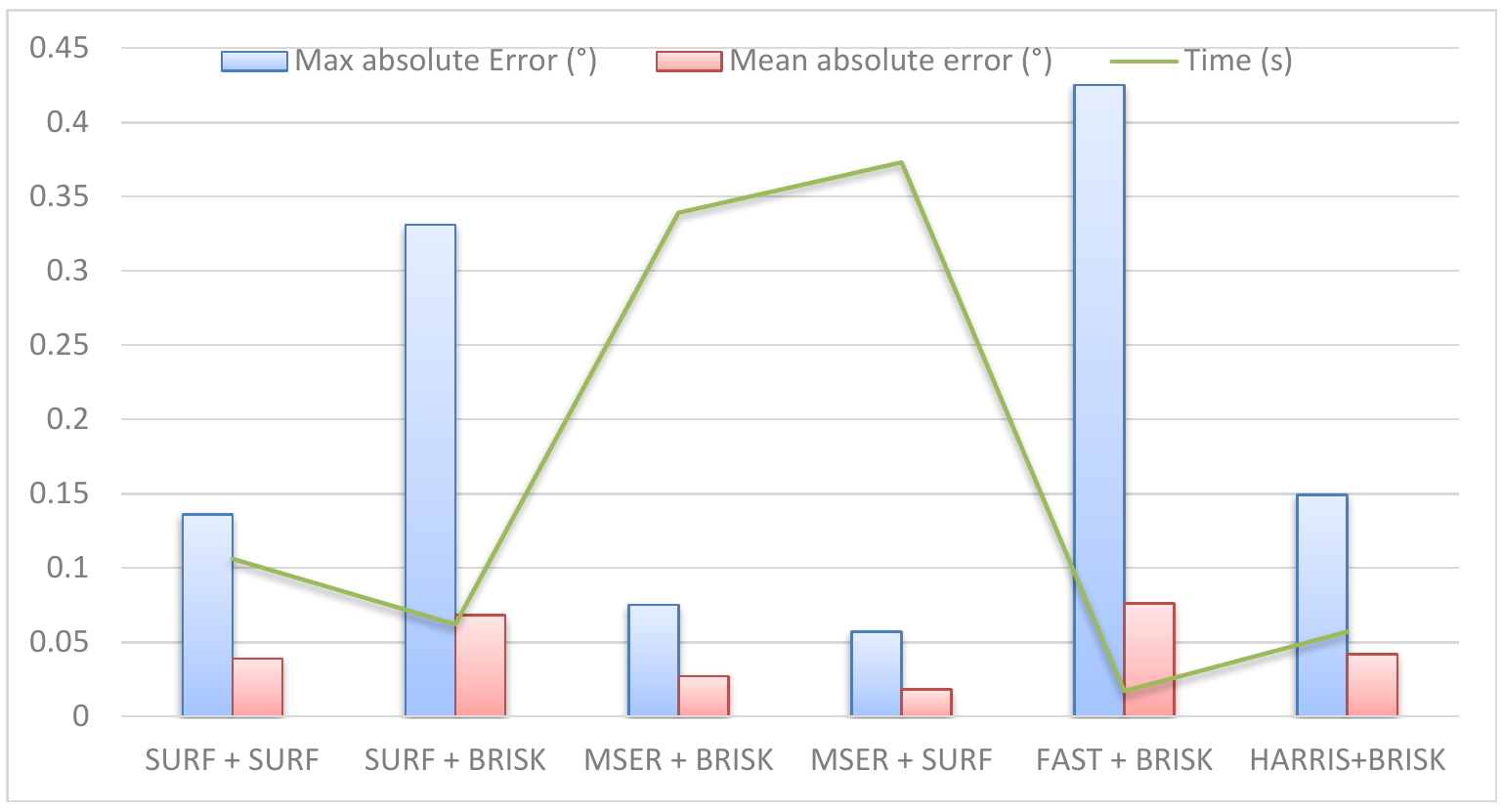}}
    \end{minipage}
\caption{Performance evaluation of rotation estimation.}
\label{fig:results_rotation}
\end{figure}

\textbf{\textit{Rotation estimation}}: We applied a similar experimental setup to that in translation estimation into the evaluation of rotation estimation. By rotating the micro stage from 0$^{\circ}$ to 10$^{\circ}$ at intervals of 1/6$^{\circ}$, we obtained 61 images and chose the first as a reference. The actual rotation angles between the test images and the reference image are successively $1/6^{\circ},1/3^{\circ},\cdots,10^{\circ}$. To evaluate the tracking of the rotation angles on tracking accuracy and computation time, we tested the same six feature extraction schemes as those tested in translation estimation. Since we simultaneously estimated translation and rotation parameters, the curves that exhibit computational cost in Figs.~\ref{fig:results_translation} and~\ref{fig:results_rotation} present a similar shape and trend. Compared with other feature extraction methods shown in Fig.~\ref{fig:results_rotation}, ``MSER+BRISK'' and ``MSER+SURF'' achieve superior tracking accuracy while sacrificing computational efficiency. Their mean absolute errors are 0.026$^\circ$ and 0.018$^\circ$, respectively, and their maximum absolute errors are 0.075$^\circ$ and 0.057$^\circ$, respectively. Among all the comparison methods, ``FAST+BRISK'' consumes the least computation time (i.e., 0.017 seconds) but yields the greatest error. In addition, by comparing feature detectors with the same feature extraction approach, we observe that MSER extracts higher quality and a larger number of blob features than SURF. Combined with the same feature detector, BRISK consumes less time than SURF. By involving tracking accuracy and computational efficiency, we choose ``MSER+BRISK" for the thread-counting system.


\textbf{\textit{Thread counting}}: The outcomes of thread counting include the basis vectors of a final lattice proposal in a camera space and two translation offsets in a lattice-based coordinate system. The mean error of translation estimation of a thread is about 0.02 (i.e., 1 thread = 0.33 mm).

\textbf{\textit{Computational time}}: It takes three frames per second (fps) for the proposed system (Matlab code without optimization, will be available online\footnotemark[2]) to calculate for a pair of images including motion estimation and thread counting. In comparison, using the algorithm from Hays et al.~\cite{Doe:2009:Online}, it takes 1.8 minutes for only lattice detection for an image of the same size. With C++ implementation (MSER and BRISK are both available in the OpenCV library) and code optimization, our proposed system is expected to operate in real time.


\begin{table}[t]
    \begin{center}
    \caption{Comparison: our system versus Book et al.~\cite{book2010automated}.}
    \label{tab:comparision}
    \resizebox{0.5\textwidth}{!}{
        \begin{tabular}{|c|c|c|c|c|c|c|}
        \hline
             & TmaxAE (pixel) & TmeanAE (pixel) & RmaxAE ($^\circ$) & RmeanAE ($^\circ$) & Time (s) & Thread\\
          \hline
         \cite{book2010automated} &  0.474  & 0.124 & 0.637 & 0.064 & 0.156 & N/A \\
          \hline
           Ours & 0.636 &    0.118    &   0.075  &  0.026&  0.339 &   Yes   \\
          \hline
        \end{tabular}
    }
    \end{center}
\end{table}

\textbf{\textit{Comparison with existing system}}: The vision system proposed by Book et al.~\cite{book2010automated} is the first and the only existing system for automatic garment sewing. Therefore, we compare the performance of their system and ours in Table~\ref{tab:comparision}, where TmaxAE, TmeanAE, RmaxAE, and RmeanAE represent translation maximum-, translation mean-, rotation maximum-, and rotation mean absolute errors, respectively. Our system outperforms theirs in two aspects: (1) our rotation tracking errors are significantly lower; and (2) our system performs thread counting, which is critical for a practical setting, but theirs cannot.

\vspace{-0.15in}
\section{Conclusion}
\label{conclusion}

We proposed an innovative thread-based texture tracking system that accurately tracks texture and detects lattice underlying fabric weave patterns in high speed. We adopted a feature extraction approach that not only detects feature points to establish valid matching between images, but also facilitates the generation of template proposal and the discovery of a repetitive lattice. To detect a reliable local lattice, we designed a computationally efficient algorithm utilizing both local appearance similarities and global topological relationship. We applied the system successfully to denim fabric tracking and thread counting, demonstrating its high potential for automatic real-time textile manufacturing.


%





%
%

\ifCLASSOPTIONcaptionsoff
  \newpage
\fi

\clearpage



%
%
%

\bibliographystyle{IEEEbib}
\bibliography{refs}

%








\end{document}